\documentclass[cameraready]{Interspeech}

\title{Benchmarking LLMs on the Massive Sound Embedding Benchmark (MSEB)}

\author[affiliation={},equalcontribution]{The MSEB}{Team}

\address{
    Google USA \& Germany
}

\email{variani@google.com}


\usepackage{comment}
\usepackage{tabularx}
\usepackage{tabularray}
\usepackage{overpic}

\begin{document}

\maketitle

\begin{abstract}

The Massive Sound Embedding Benchmark (MSEB) has emerged as a standard for evaluating the functional breadth of audio models. While initial baselines focused on specialized encoders, the shift toward "audio-native" Large Language Models (LLMs) suggests a new paradigm where a single multimodal backbone may replace complex, task-specific pipelines. This paper provides a rigorous empirical evaluation of leading LLMs - including members from the Gemini and GPT families - across the eight core MSEB capabilities to assess their efficacy and audio-text parity. 
Our results indicate that while a significant modality gap persists regarding performance and robustness, the empirical evidence for an "optimal" modeling approach remains inconclusive. Ultimately, the choice between audio-native and cascaded architectures depends heavily on specific use-case requirements and the underlying assumptions regarding latency, cost, and reasoning depth.
\end{abstract}

\section{Introduction}\label{sec:intro}

The landscape of artificial intelligence is currently undergoing a paradigm shift, transitioning from specialized, unimodal systems toward integrated auditory intelligence. This evolution is primarily driven by the emergence of "Audio-Native" Large Language Models (LLMs). 
Unlike previous generations that relied on cascaded pipelines—where standalone audio encoders like Whisper served as a front-end for a text-based LLM—state-of-the-art models such as Gemini 3 and GPT-5 integrate audio processing directly into their neural fabric. 
These multimodal architectures aim to replace traditional, task-specific encoders with a unified system capable of sophisticated, end-to-end auditory reasoning. 
As the industry moves toward this audio-native standard, there is a critical need for a rigorous framework to measure progress across the full spectrum of sound understanding.

The recently introduced Massive Sound Embedding Benchmark (MSEB)~\cite{google2025mseb} provides a comprehensive suite of tasks designed to capture the core requirements of auditory intelligence, making it uniquely suited to evaluate these next-generation LLMs. MSEB is characterized by its breadth; it is inherently multilingual—spanning both high- and low-resource languages - and covers a diverse range of acoustic environments. Specifically, MSEB evaluates models across eight distinct "supertasks":
\begin{itemize}
    \item Transcription: Traditional and multilingual speech-to-text.
    \item Retrieval: Voice search and knowledge base querying.
    \item Reasoning: Direct question answering based on audio input.
    \item Classification: Identifying intent, environments, and specific sound events.
    \item Reranking: Refining transcription hypotheses for higher accuracy.
    \item Segmentation: Audio indexing and precise timestamp localization.
    \item Clustering: Unsupervised grouping of diverse audio samples.
    \item Reconstruction: Measuring embedding fidelity through waveform regeneration.
\end{itemize}

This paper presents a rigorous empirical evaluation of leading LLMs on the MSEB benchmark to quantify their current auditory capabilities and assess audio-text parity. Our primary contributions are as follows:
\begin{itemize}
    \item We adapt multimodal LLMs to solve a diverse range of MSEB audio tasks - encompassing both generative and non-generative challenges - via task-specific prompting (Section~\ref{sec:methodology}).
    \item We perform a comprehensive evaluation (Section~\ref{sec:evaluation}) comparing modern multilingual LLMs against state-of-the-art specialized and cascaded systems.
    \item We analyze audio-text parity across multiple dimensions, including performance headroom, architectural modeling approaches, acoustic robustness, and locale-specific variance (Section~\ref{sec:audio-text-parity}).
    \item We contribute our results to the public leaderboard and open-source the implementation by extending the MSEB toolkit\footnote{\url{https://github.com/google-research/mseb}}.
\end{itemize}

The remainder of this paper is organized as follows: Section~\ref{sec:methodology} details the application of LLMs to MSEB tasks, including the specific prompt templates utilized. Section~\ref{sec:setup} describes the experimental datasets and models. Section~\ref{sec:evaluation} presents our experimental results and analysis, followed by a discussion on audio-text parity across several dimensions in Section~\ref{sec:audio-text-parity}. Finally, concluding remarks are presented in Section~\ref{sec:conclusion}.

\section{Related Work}\label{sec:related}

The evolution of auditory intelligence has transitioned from modular, task-specific pipelines toward unified, audio-native architectures. This section situates our evaluation within the broader context of recent architectural shifts and the benchmarks developed to assess them.


A new generation of audio-native multimodal LLMs has emerged. Examples include recent versions of LLM families such as Gemini~\cite{gemini2025audio}, GPT~\cite{openai2024gpt4o}, Gemma~\cite{gemma20253n}, Qwen~\cite{xu2025qwen25omnitechnicalreport} integrate audio features directly into a shared latent space, allowing for end-to-end reasoning without intermediate text bottlenecks. Similarly, embeddings models with audio support have been published recently, including Amazon Nova~\cite{amazon2025nova} and Gemini embedding V3. While an active field with promising results, true parity with text-based reasoning remains an open research challenge~\cite{ma2025mmar}.

As model capabilities have expanded, so too has the scope of evaluation frameworks. Early benchmarks like SUPERB~\cite{Yang2021SUPERBSP} focus on classical speech tasks such as speaker identification and phoneme recognition. Dynamic-SUPERB~\cite{Huang2023DynamicSuperbTA} is a comprehensive, multi-task English benchmark for evaluating speech processing models in zero-shot scenarios, scaling SUPERB from 33 to 180 tasks. More recently, AudioBench~\cite{wang-etal-2025-audiobench} introduced a universal benchmark designed to evaluate Audio LLMs, with a focus on understanding English semantics. MMAU~\cite{Sakshi2024MMAUAM} and MMAU-Pro~\cite{kumar2025mmau-pro} target complex, multi-hop reasoning for expert-level auditory analysis in English. Possibly the most recent developments are the Massive Audio Embedding Benchmark (MAEB), which builds upon the MTEB ecosystem to cover diverse domains like music and bioacoustics across 100+ languages, and Google’s Massive Sound Embedding Benchmark (MSEB). Unlike the broader scope of MAEB~\cite{assadi2026maebmassiveaudioembedding}, MSEB~\cite{google2025mseb} is centered around real-world multimodal speech tasks across 25+ locales, including long-context document retrieval.

\section{Methodology: Applying LLMs to MSEB}\label{sec:methodology}

This section details the technical methodology for addressing the various tasks within the MSEB benchmark using LLMs. While LLMs are natively suited for generative tasks—such as reasoning - adapting them to non-generative tasks like retrieval remains challenging. 
To maintain a unified framework, we integrate these diverse tasks by standardizing them through a task-specific prompt template:
\vspace{1mm}\fbox{
\begin{minipage}{0.45\textwidth}
\addtolength{\leftskip}{0mm} 
\vspace{1mm}

\textbf{Task:} Name of task.

\textbf{Goal:} High-level task description.

\textbf{Input:} Description of the input(s) and the expected format.
 
\textbf{Output:} Description of the expected output and the format.

\textbf{Important Considerations:} Additional constraints etc.

\texttt{\{\{"query": \{text\}, etc.\}\}}

\vspace{1mm}
\end{minipage}
}\vspace{1mm}
The initial prompt template was refined iteratively through interaction with Gemini 3 until the outputs consistently met the required structured format.
In the remainder of this section, we define the prompt engineering strategy for each supertask in detail, roughly sorted from most generative to least (i.e., discriminative) in nature.


\paragraph*{Reasoning:}
This task involves answering spoken questions based exclusively on a provided source passage. 
Since not all queries can be resolved using the given text, the model must also assess answerability to prevent hallucinations. 
The input consists of the source text passage alongside the query, provided either as a wav-file or a text transcript. 
The output is a structured JSON string: it contains an 'answer' key if the information is present, or a 'no\_answer' key if the context is insufficient.

\paragraph*{Speech transcription:} 
The task of speech transcription is a native capability of LLMs and provides a verbatim transcription of the provided audio clip.
The audio with the speech is passed as a wav-file. The output is a plain text string containing the transcription.


\paragraph*{Salient term segmentation:}
This task requires identifying and localizing the most informative keywords within an audio query. For this task, a "salient term" is defined as a single-word concept or topic that is highly specific — statistically likely to appear in only a small subset of Wikipedia articles. The objective is to extract the top-$k$ most relevant terms, providing for each a structured JSON object containing the term itself along with its precise start and end times in seconds. To ensure compatibility with automated downstream processing, the model must output these objects as raw, plain-text strings (one per line) without any Markdown formatting or code blocks.

\paragraph*{Query reranking:}
Query reranking serves as a second-pass refinement of transcription hypotheses. 
The model receives the spoken query as a wav-file or transcript alongside a set of acoustically and semantically similar candidates, provided as a structured JSON string. 
The output is a permutation of this list, sorted by the predicted accuracy of the candidate transcripts.

\paragraph*{Classification:}
In the classification task, we assign a label from a predefined finite set to an audio sample. 
LLMs are particularly well-suited for zero-shot classification. 
To perform this, we provide the model with the audio file (e.g., in WAV format) or a transcript of it alongside a list of descriptive class labels. 
The model is then tasked with returning the most appropriate label from the provided set. 
Additionally, non-answerability is supported, allowing the model to indicate if no suitable label exists.

\paragraph*{Document retrieval:} 
Retrieval in this framework entails identifying the most relevant entries from a document (Wikipedia passages or pages) corpus based on a spoken query (e.g., voice-activated search). 
Since including the entire index in the prompt would exceed context window constraints, we employ a Retrieval-Augmented Generation (RAG) pipeline. 
In this setup, an external embedding model performs an initial retrieval of the top-$k$ candidates. 
We utilize a dynamic $k$ strategy, where the number of retrieved candidates is adjusted to fit into the LLM's context window. 
The LLM then acts as a \textbf{re-ranker}, sorting these candidates by their relevance to the query and returning the results in a structured JSON format.


\section{Experimental Setup}\label{sec:setup}

In this section, we briefly describe the models and the Massive Sound Embedding Benchmark (MSEB)~\cite{google2025mseb} utilized for the empirical evaluation and analysis presented in Section~\ref{sec:evaluation}.

\subsection{Models Evaluated}

Our evaluation covers a diverse set of audio-native MLLMs, including both commercially restricted (proprietary) and open-weight models, selected for their competitive performance on auditory reasoning tasks.


\paragraph*{Gemini 2.5 \& 3 Flash\footnote{\url{https://deepmind.google/models/gemini/flash/}}:}
These proprietary, high-throughput multimodal models employ a highly optimized Mixture-of-Experts (MoE) architecture, totaling approximately 128B parameters. They are engineered specifically for low-latency reasoning and high-concurrency applications~\cite{allbert2025evaluating}. 
We utilized these models due to their native support for interleaved text, audio, and video modalities, their industry-leading 1-million-token context window, and their specialized distilled reasoning engine, which provides near-Pro-level accuracy on retrieval tasks at a fraction of the computational cost.



\paragraph*{GPT-4o-mini / GPT-4o-mini-audio\footnote{\url{https://openai.com/index/gpt-4o-mini-advancing-cost-efficient-intelligence/}}:}
A high-efficiency proprietary multimodal model from OpenAI, specifically designed to balance low latency with sophisticated multimodal reasoning. GPT-4o-mini-audio features a 128,000-token context window, optimized for the cost-effective processing of diverse data streams without compromising retrieval accuracy. Native multimodality is a foundational element of its architecture, enabling the unified and end-to-end processing of text, image, and audio inputs within a single transformer-based framework. 
For text-only ablations involving transcriptions without audio input, we employ GPT-4o-mini as an equivalent model.


In addition to LLMs, we employ Gemini embedding as the dense embedding model to facilitate retrieval tasks within our RAG-based framework. 

\paragraph*{Gemini embedding\footnote{\url{https://ai.google.dev/gemini-api/docs/embeddings}}:}
For semantic retrieval, we utilize Gemini Embedding, a proprietary, high-dimensionality dense vector model. Initialized from the Gemini transformer backbone, the model supports a context window of 2,048 tokens and produces embeddings with a native dimensionality of 3,072. 
While primarily optimized for text and code, its pre-training on the multimodal Gemini corpus ensures robust semantic representation across 100+ languages~\cite{lee2025geminiembeddinggeneralizableembeddings}. 


Finally, we employ several Automatic Speech Recognition (ASR) front-ends to convert audio input into text for use in the cascaded approach. This textual conversion is also necessary to interface with our current RAG pipeline, which utilizes text-based Gemini embeddings for document retrieval.

\paragraph*{Whisper large,v3\footnote{\url{https://huggingface.co/openai/whisper-large-v3}} \& GPT-4o-transcribe\footnote{\url{https://platform.openai.com/docs/guides/speech-to-text}}:}
We leverage two flagship automatic speech recognition (ASR) systems to provide high-fidelity textual representations for our cascaded pipelines. Whisper large,v3 serves as our primary open-source baseline; it is a Transformer-based encoder-decoder model trained on 680,000 hours of multilingual data, utilizing 128 Mel frequency bins to achieve superior robustness in noisy acoustic environments. Complementing this, we utilize GPT-4o-transcribe, OpenAI’s state-of-the-art proprietary ASR engine. Unlike previous iterations that relied on separate pipelines, GPT-4o-transcribe leverages the native audio-reasoning capabilities of the GPT-4o architecture, providing significant reductions in Word Error Rate (WER) and enhanced sensitivity to prosodic nuances and diverse linguistic locales. Together, these models ensure that our textual front-end captures the highest possible semantic density before further processing.

\paragraph*{ElevenLabs\footnote{\url{https://elevenlabs.io/}}}
ElevenLabs, primarily known for its industry-leading generative voice AI, also offers a high-performance speech-to-text model designed for speed and accuracy.
Unlike traditional ASR models that struggle with tone and context, ElevenLabs leverages its deep understanding of prosody and linguistics to provide transcriptions that are highly resilient to accents and background noise. It is often positioned as a premium alternative to OpenAI's Whisper, optimized for low-latency applications like real-time captioning, conversational AI, and media post-production.

\subsection{Massive Sound Embedding Benchmark (MSEB)}
MSEB includes a collection of public datasets and an open-source toolkit.

\paragraph*{Datasets:}
MSEB comes with a diverse suite of datasets that span linguistic, environmental, and bioacoustic domains. 
\textbf{Simple Voice Questions (SVQ)\footnote{\url{https://huggingface.co/datasets/google/svq}}} provides a multilingual foundation for several tasks, consisting of over 177,000 spoken queries with metadata in 17 languages and 26 locales recorded across varying noise conditions.
\textbf{Speech-MASSIVE} extends the well-known MASSIVE textual corpus into the auditory realm, offering a multilingual benchmark for spoken language understanding (SLU) with intent and slot annotations across 12 languages~\cite{speech-massive}. 
For environmental sound recognition, we employ \textbf{FSD50K}, a large-scale, open dataset containing over 51,000 human-labeled sound events across 200 classes from the AudioSet Ontology~\cite{fsd50k}. 

\paragraph*{Toolkit:}
All evaluations were conducted using the open-source MSEB toolkit\footnote{\url{https://github.com/google-research/mseb}}. 
The framework provides standardized task definitions, automated data loading, and modular wrappers for LLM encoders. Performance is quantified using task-specific metrics: top-1 accuracy (classification), V-measure (clustering), MRR@10 (retrieval), MAP (reranking), Word Error Rate (WER) (speech transcription), F1-score (reasoning), NDCG (segmentation), and LSD (reconstruction).
\section{Empirical Evaluation}\label{sec:evaluation}


We present and discuss the results on a task-by-task basis in the following subsections. A high-level overview of these results is provided in Table~\ref{tab:llm-by-task}. Furthermore, Section~\ref{sec:audio-text-parity} offers a more granular analysis of these findings through the lens of audio-text parity.

\begin{table*}[ht]
    \centering
    \caption{Model capability comparison across MSEB tasks. Results reported on the SVQ dataset for all tasks except for the intent and sound classification tasks. 'Fail' indicates that no prompt was found that could reliably elicit the target task behavior.}
    \label{tab:llm-by-task}
    \vspace{-2ex}
    \footnotesize
    \begin{tabularx}{\textwidth}{l|p{1.2cm}p{1.2cm}p{1.2cm}p{1.2cm}p{1.2cm}p{1.2cm}p{1.2cm}p{1.2cm}p{1.2cm}p{1.2cm}}
        \toprule
        \textbf{Model} & \textbf{Speech} & \textbf{Query} & \textbf{Salient} & \textbf{Intent} & \textbf{Gender} & \textbf{Sound} & \textbf{Passage} & \textbf{Passage} & \textbf{In-Lang} \\
        \textbf{\textit{\ \ Input\ modality}} & \textbf{Transcr.} (WER$\downarrow$) & \textbf{Rerank} (MAP$\uparrow$) & \textbf{Term Segment.} (NDCG$\uparrow$) & \textbf{Classif.} (Acc.$\uparrow$) & \textbf{Classif.} (Acc.$\uparrow$) & \textbf{Classif.} (mAP$\uparrow$) & \textbf{In-Lang Retrieval} (MRR$\uparrow$) & \textbf{Cross-Lang Retrieval} (MRR$\uparrow$) & \textbf{Reason.} (F1$\uparrow$) \\
        \midrule
        Whisper & .329~\cite{google2025mseb} & & & & & & & & \\
        ElevenLabs & .288 & & & & & & & & \\
        GPT-4o-transcribe & .249 & & & & & & & & \\
        \hline
        LAION Clap & & & & & .910 & .432 & & & \\
        \hline
        Gemini embedding & & & & & & & & & \\
        \ \ \textit{Whisper} & & .599~\cite{google2025mseb} & & .803~\cite{google2025mseb} & & & .619~\cite{google2025mseb} & .502~\cite{google2025mseb} & \\
        \ \ \textit{GPT-4o-transcribe} & & & & & & & .641 & .569 & \\
        \ \ \textit{Truth} & & 1.00 & & .834~\cite{google2025mseb} & & & .759~\cite{google2025mseb} & .683~\cite{google2025mseb} & \\
        \hline
        GPT-4o-mini-audio & & & & & & & & & \\
        \ \ \textit{Audio} & fail & .522 & & .598 & fail & fail & .609 & .587 & (.490) \\
        \ \ \textit{GPT-4o-transcribe} & & & & & & & .608 & .587 & \\
        \ \ \textit{Truth} & & & & & & & .689 & .688 & \\
        Gemini 2.5 Flash &  & & & & & & & & \\
        \ \ \textit{Audio} & .355 & & .345 & .752 & .871 & .120 & (.697) & (.536) & .558 \\
        \ \ \textit{Whisper} & & & .351 & .776 & & & .662 & (.567) & .587 \\
        \ \ \textit{Truth} & & & & .805 & & & .798 & (.724) & .581 \\
        Gemini 3 Flash & & & & & & & & & \\
        \ \ \textit{Audio} & .311 & & & .801 & .880 & .218 & & & .588 \\
        \ \ \textit{GPT-4o-transcribe} & & & & & & & & & \\
        \ \ \textit{Truth} & & & & .836 & & & & & \\
        \bottomrule
    \end{tabularx}
\end{table*}

\subsection{Speech transcription}
\begin{figure}[t]
  \centering
    \includegraphics[width=\linewidth]{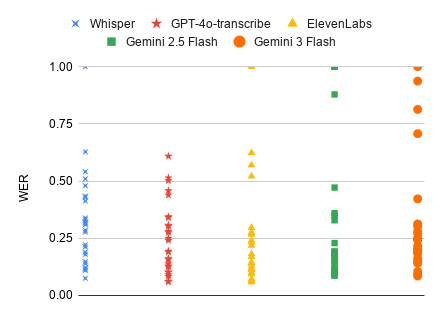}
  \vspace{-2ex}
  \caption{Speech transcription (WER) for different custom (Whisper, GPT-4o-transcribe, ElevenLabs) and LLM (Gemini 2.5 / 3 Flash) models. Each marker represents a model/locale pair.}
  \label{fig:wer-scatter}
\end{figure}

WER averaged across all locales for various custom and audio-native LLMs is presented in Table~\ref{tab:llm-by-task}. Notably, GPT-4o-mini-audio exhibited significant instruction-following challenges; we were unable to identify a prompt that consistently yielded verbatim transcripts without the model introducing conversational commentary, answering embedded questions, or refusing the request. 
GPT-4o-transcribe is the top-performing model by a clear margin.
A breakdown by locale provides a more nuanced perspective (see Fig.~\ref{fig:wer-scatter}). The variance across locales is substantial; notably, every model except GPT-4o-transcribe produced outliers exceeding 100\% WER in certain locales. These extreme values significantly affect the aggregate average WER.

\subsection{Query reranking}
\begin{figure}[t]
  \centering
    \includegraphics[width=\linewidth]{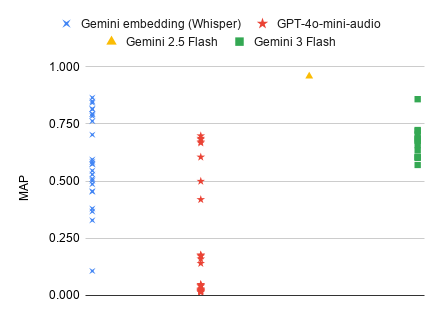}
  \vspace{-2ex}
  \caption{Query reranking (MAP) for different an embedding (Gemini embedding) and various LLM (GPT-4o-mini-audio, Gemini 2.5 / 3 Flash) models. Each marker represents a model/locale pair.}
      \label{fig:query-reranking-scatter}
\end{figure}

MAP across all locales for various embedding models and audio-native LLMs is presented in Table~\ref{tab:llm-by-task}, with a detailed locale-wise breakdown in Fig.~\ref{fig:query-reranking-scatter}. The variance across locales remains substantial.
While GPT-4o-mini-audio does not refuse query reranking requests, its performance is only moderate and lags behind the cascaded embedding approach; in contrast, initial MAP figures for Gemini 2.5 Flash and Gemini 3 Flash are significantly higher.
For perspective, the lower bound established by random permutations of candidates ranges between 0.02 and 0.03.

\subsection{Salient term segmentation}

Table~\ref{tab:segmentation} compares the performance of some ASR and Gemini LLMs models in the salient term segmentation task. The GPT-4o Transcibe models where excluded from the set of ASR models considered since they cannot produce word timings.

On this task, the ASR-based models are at an advantage since they have access to the IDF table used to create the reference salient terms. Whisper even more so since the reference segment timings were obtained by force alignment using that model.

ElevenLabs achieves the highest scores for NDCG and term accuracy, narrowly outperforming Whisper. The Gemini models produce mediocre term accuracy results leading to poor NDCG. Expected, the segment timing accuracy is abyssal.

\begin{table}[ht]
\centering
\caption{Salient term segmentation results}
\label{tab:segmentation}
\vspace{-2ex}
\small
\begin{tabularx}{\columnwidth}{l|ccc}
\toprule
\textbf{Model} & \textbf{NDCG} & \textbf{Term Acc.} & \textbf{Timing Acc.}\\
\midrule
Whisper & .433 & .549 &  .407\\ 
ElevenLabs & .608 & .718 & .135\\
Gemini 2.5 Flash & .222 & .392 & .017\\
Gemini 3 Flash & .255 & .430 & .007\\
\bottomrule
\end{tabularx}
\end{table}
\subsection{Intent classification}

\begin{figure}[t]
  \centering
    \includegraphics[width=\linewidth]{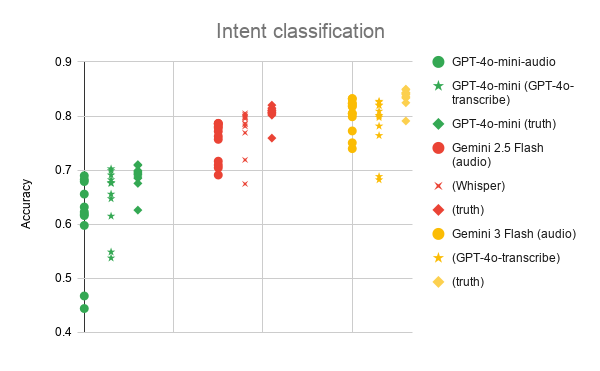}
  \vspace{-2ex}
  \caption{Intent classification accuracy across architectures. Each marker represents a specific model-locale configuration, highlighting the performance stability (or lack thereof) across diverse languages.}
      \label{fig:intent-classification-scatter}
\end{figure}

Table~\ref{tab:intent_classification} shows a strong performance from the Gemini models (both embeddings and generative LLMs). 
The locale-specific breakdown in Fig.~\ref{fig:intent-classification-scatter} corroborates these high scores
However, our interaction with Gemini 2.5 Flash showed that the model is aware of the SLURP/MASSIVE datasets and recognize the class labels as coming from there. This probably explains why these models outperform the GPT-4o (mini) Audio models, see also Section~\ref{sec:data_contamination}.


\begin{table}[ht]
\centering
\caption{Intent classification accuracy}
\label{tab:intent_classification}
\vspace{-2ex}
\small
\begin{tabularx}{\columnwidth}{l|ccl}
\toprule
\textbf{Model} & \textbf{Audio} & \textbf{Truth} & \textbf{ASR}\\
\midrule
Gemini Embeddings &  &  .834 & .803 {\footnotesize(Whisp.)} \\
\midrule
Gemini 2.5 Flash & .752 & .805 &  .776 {\footnotesize(Whisp.)} \\
Gemini 2.5 Pro & .774 & .812 &  .783 {\footnotesize(Whisp.)} \\
Gemini 3.0 Flash & .801 & .836 & .785 {\footnotesize(GPT-4o Tr.)} \\
GPT-4o Mini (Audio) & .598 & .689 & .651 {\footnotesize(GPT-4o Tr.)} \\
GPT-4o (Audio) & .683 & .780 & \\
\bottomrule
\end{tabularx}
\end{table}

\subsection{Speaker-gender classification}

Table~\ref{tab:gender_classification} presents accuracy on both the SVQ and Speech-MASSIVE datasets. On the SVQ dataset, LAOIN Clap strongly outperforms the Gemini models despite its much smaller size and simpler architecture. The Speech-MASSIVE dataset presents an easier challenge and the larger Gemini models do outperform Clap there. 
We don't know if the Gemini models have seen the Speech-MASSIVE during training.
The GPT-4o (mini) Audio models refused to perform the task claiming they cannot perform speaker classification.

\begin{table}[ht]
\centering
\caption{Speaker-gender classification accuracy.}
\label{tab:gender_classification}
\vspace{-2ex}
\small
\begin{tabularx}{\columnwidth}{l|ccc}
\toprule
\textbf{Model} & \textbf{SpeechMassive} & \textbf{SVQ} \\
\midrule
LAION Clap & .952 &  .910\\
\midrule
Gemini 2.5 Flash Lite & .860 & .767\\
Gemini 2.5 Flash & .961 & .871\\
Gemini 2.5 Pro & .976 & \\
Gemini 3 Flash &  & .880\\
\bottomrule
\end{tabularx}
\end{table}

\subsection{Sound classification}
Table~\ref{tab:sound_classification} shows that LLMs are not competitive on this task with the much smaller LAOIN Clap vector-embedding model as far as precision is concerned. The GPT-4o (mini) Audio models refused to perform this task due to a claimed lack of competency. The larger Gemini models do perform better than their smaller counterpart and can even achieve better F1 scores than Clap, compensating their weak precision with strong recall.

\begin{table}[ht]
\centering
\caption{Sound classification performance on FSD50k.}
\label{tab:sound_classification}
\vspace{-2ex}
\small
\begin{tabularx}{\columnwidth}{l|ccc}
\toprule
\textbf{Model} & \textbf{mAP} & \textbf{Micro F1} & \textbf{Macro F1} \\
\midrule
LAION Clap & 0.432 &  0.183 &  0.265 \\
\midrule
Gemini 2.5 Flash Lite & 0.058 & 0.186 & 0.144\\
Gemini 2.5 Flash & 0.120 & 0.288 & 0.246 \\
Gemini 2.5 Pro & 0.245 & 0.477 & 0.429\\
Gemini 3 Flash & 0.218 & 0.432 & 0.388\\
\bottomrule
\end{tabularx}
\end{table}

\subsection{Retrieval} 
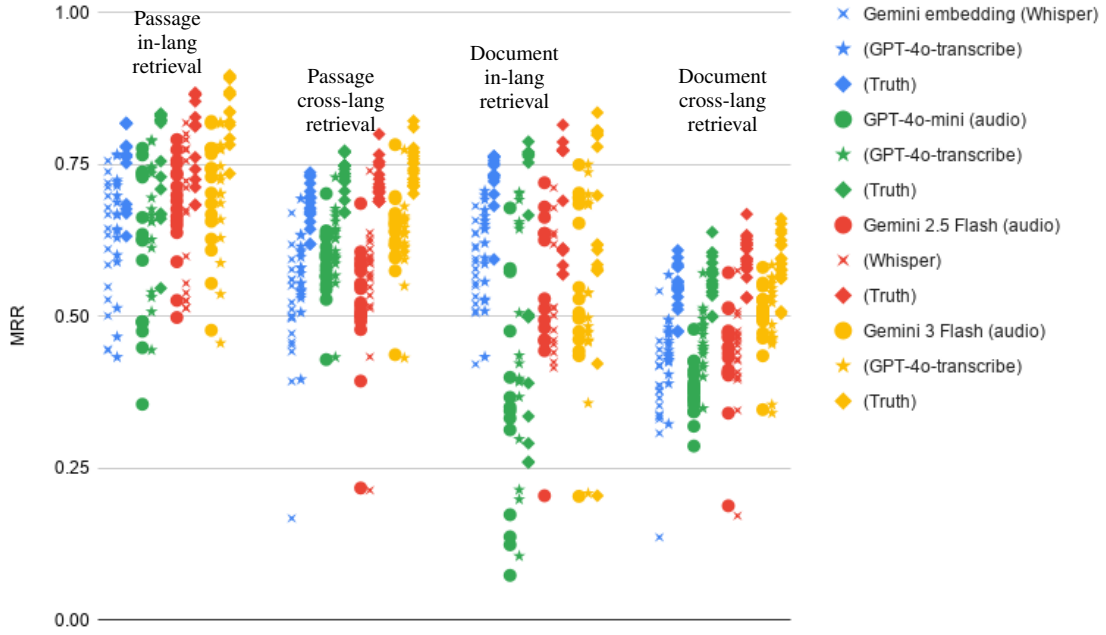
\begin{figure*}[t]
 \centering
\begin{overpic}[width=0.9\textwidth]
{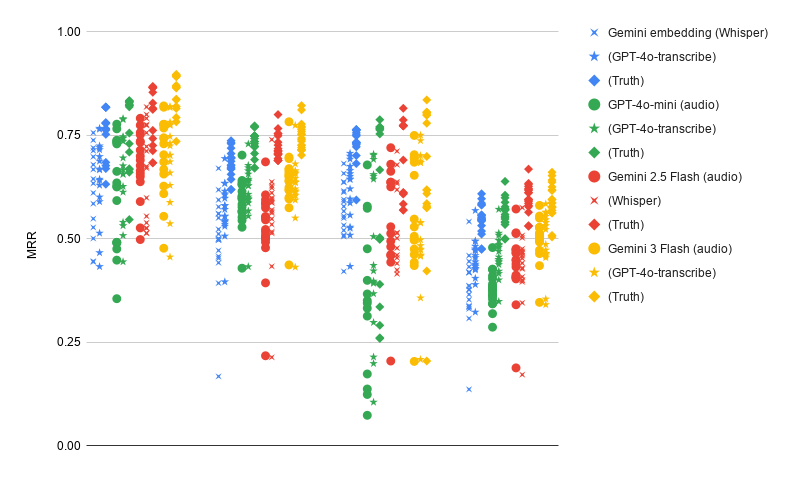}
\put(12,55){\parbox{1.5cm}{\footnotesize\centering Passage in-lang retrieval}}
\put(27,50){\parbox{1.5cm}{\footnotesize\centering Passage cross-lang retrieval}}
\put(42,52){\parbox{1.5cm}{\footnotesize\centering Document in-lang retrieval}}
\put(60,50){\parbox{1.5cm}{\footnotesize\centering Document cross-lang retrieval}}
\end{overpic}
\vspace{-2ex}
\caption{MRR performance for the four retrieval variants across various models. Each marker represents a unique model-locale pair, illustrating the variance in retrieval effectiveness across locales.}
\label{fig:retrieval}
\end{figure*}

Fig. \ref{fig:retrieval} compares the performance of the four retrieval variants. Notably, the Gemini embedding is the only model performing standalone retrieval, whereas the other models act as re-rankers within a RAG pipeline, scoring the top-$k$ items initially retrieved by the Gemini embedding (see Table~\ref{tab:recall_at_10} for Recall@10).

The models ranked by MRR in increasing order are: Gemini embedding, GPT-4o-transcribe, Gemini 2.5 Flash, and Gemini 3 Flash. 
Excluding outliers caused by limited coverage in specific locales, a comparison between cascaded and native audio reveals no clear performance advantage for either approach. Notably, the substantial reductions in WER achieved by GPT-4o-transcribe over Whisper do not translate into corresponding improvements in MRR.
See Section~\ref{sec:native_vs_cascaded_audio} for further discussion.

\begin{table}[ht]
    \centering
    \caption{Recall@10 for RAG-based tasks.}
    \label{tab:recall_at_10}
    \vspace{-2ex}
    \small
    \begin{tabularx}{\columnwidth}{l|XXXX}
        \toprule
        \textbf{Model} & \multicolumn{2}{c}{\textbf{Passage}} & \multicolumn{2}{c}{\textbf{Document}} \\
        & \textbf{In-Lang} & \textbf{Cross-Lang} & \textbf{In-Lang} & \textbf{Cross-Lang} \\
        \midrule
        Whisper & .801 & .710 & (.724) & (.548) \\
        GPT-4o-transcribe & .817 & .776 & (.745) & (.645) \\
        Truth & .938 & .905 & (.891) & (.779) \\
        \bottomrule
    \end{tabularx}
\end{table}

\subsection{Reasoning}

\begin{figure}[t]
  \centering
    \includegraphics[width=\linewidth]{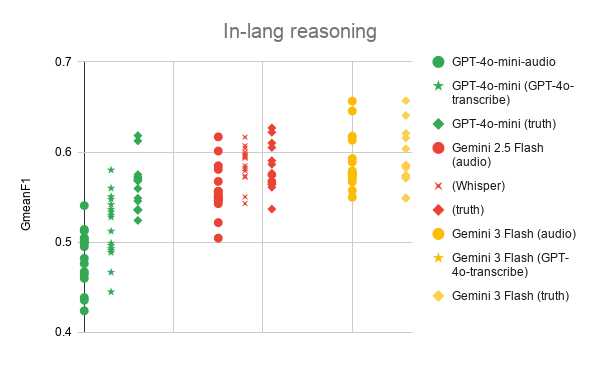}
  \vspace{-2ex}
  \caption{In-language reasoning performance across models. Each marker represents the GmeanF1 score for a specific model-locale pair, illustrating the variance in reasoning capabilities across different languages.}
      \label{fig:reasoning-scatter}
\end{figure}

Gemini 3 Flash is the best performing model for the in-language span-reasoning tasks, as shown by the GmeanF1 results in Table~\ref{tab:reasoning}. This model also appears to have effectively closed the modality gap, as the aggregate results from audio input closely match those obtained from the reference transcripts. Fig.~\ref{fig:reasoning-scatter} provides a further breakdown by locale for additional insight.

\begin{table}[ht]
\centering
\caption{GmeanF1 for the in-language span reasoning tasks.}
\label{tab:reasoning}
\vspace{-2ex}
\small
\begin{tabularx}{\columnwidth}{l|ccl}
\toprule
\textbf{Model} & \textbf{Audio} & \textbf{Truth} & \textbf{ASR}\\
\midrule
Gemini 2.5 Flash Lite & .477 & .561 &  .549 {\footnotesize(Whisper)} \\
Gemini 2.5 Flash & .558 & .581 &  .587 {\footnotesize(Whisper)} \\
Gemini 2.5 Pro   & .567 & .584 & \\
Gemini 3 Flash   & .588 & .589 &  \\ 
GPT-4o Mini (Audio) & .484 & .562 & .518 {\footnotesize(GPT-4o Tr.)}\\
GPT-4o (Audio)      & .555 & .589 & \\
\bottomrule
\end{tabularx}
\end{table}

\section{Audio-text parity}\label{sec:audio-text-parity}

Audio-text parity refers to the functional equivalence between processing a prompt in audio format versus its textual counterpart. Informally, this implies that the "I type" and "I speak" modalities achieve comparable performance across tasks, ensuring that no information or reasoning capability is lost when switching from text to speech.

Based on the evaluation results in Section~\ref{sec:evaluation}, we analyze this parity across several key dimensions, including headroom (speaking vs typing mode), modeling approach (audio-native vs cascaded), acoustic robustness, and evaluation cost.

\subsection{Headroom}
We define headroom as the performance gap between a model provided with the reference text versus a "noisy" transcript version. In the context of speech tasks, this represents the difference in metrics between using the ground-truth transcript and an ASR-generated transcript. 
The headroom depends on the model. 
A narrow headroom may suggest that audio-text parity has been achieved. However, in several instances, it instead indicates that the model is underperforming across both audio and text modalities (e.g., in reasoning tasks) or that test data contamination is artificially inflating scores (e.g., in intent classification; see Section~\ref{sec:data_contamination}).
We consolidate the key metrics from Section~\ref{sec:evaluation} and summarize the resulting headrooms in Table~\ref{tab:headroom}.
\begin{table}[ht]
\centering
\caption{Performance headroom across distinct tasks.}
\label{tab:headroom}
\vspace{-2ex}
\begin{tabular}{@{}ll|ccc@{}}
\toprule
\textbf{Task} & \textbf{Model} & \rotatebox{90}{\parbox{1.6cm}{\raggedright\textbf{Audio nat. (tokenized)}}} & \rotatebox{90}{\parbox{1.6cm}{\raggedright\textbf{ASR transcript}}} & \rotatebox{90}{\parbox{1.6cm}{\raggedright\textbf{Transcript truth}}} \\ 
\midrule
Sal. term & Whisper \& IDF & & .402 & \\
segment. & Gemini 2.5 Flash & .345 & .351 & \\
\hline
Intent & Gemini 2.5 Flash & .752 & .776 & .805 \\
classif. & Gemini 3 Flash & .801 & & .836 \\
\hline
Passage & Gemini embedding & & .641 & .759 \\
in-lang & GPT-4o-mini & .609 & .608 & .689 \\
retrieval & Gemini 2.5 Flash & .697 & .662 & .798 \\
\hline
Passage & Gemini embedding & & .569 & .683 \\
cross-lang & GPT-4o-mini & .587 & .587 & .688 \\
retrieval & Gemini 2.5 Flash & .536 & .567 & .724 \\
\hline
In-lang & Gemini 2.5 Flash & .558 & .587 & .580 \\
reasoning & Gemini 2.5 Pro & .567 & & .584 \\
\bottomrule
\end{tabular}
\end{table}

\subsection{Modeling approach}\label{sec:native_vs_cascaded_audio}
We compare the two prevailing modeling approaches:
\begin{itemize}
\item \textbf{Audio-native:} Multimodal LLMs that process audio directly. The audio signal is typically discretized into latent tokens, allowing the model to reason on the acoustic representation alongside text. LLMs are fine-tuned on the audio tokens.
\item \textbf{Cascaded:} A modular pipeline where an Automatic Speech Recognition (ASR) system first converts audio into text, which is then fed into a text-based LLM. Unlike audio-native models, general-purpose LLMs are typically not explicitly fine-tuned on ASR transcripts. 
\end{itemize}


The per-task evaluation in Section~\ref{sec:evaluation} suggests performance parity between cascaded and native audio for Gemini 2.5 and 3 Flash. In contrast, the native audio capabilities of GPT-4o-mini-audio underperform relative to Gemini and cascaded approach, likely due to a design focus on high-fidelity transcription paired with text-based reasoning rather than native multimodal audio processing.

\begin{figure}[t]
  \centering
    \centering
    \includegraphics[width=\linewidth]{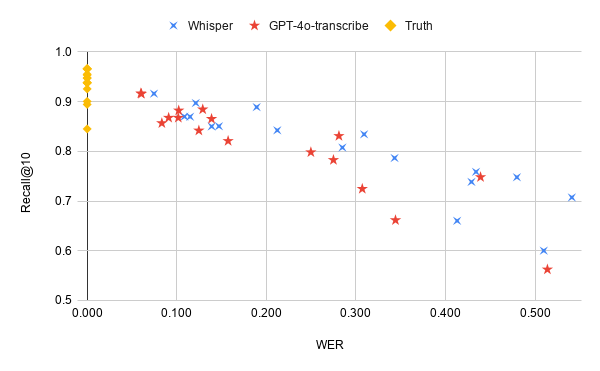}
  \vspace{-2ex}
  \caption{Recall@10 for the Gemini embedding model, utilized as the initial retriever for the RAG-based LLM variants. Results are shown for the in-language passage retrieval task.}
  \label{fig:wer-recall_at_10}
\end{figure}

\begin{figure}[t]
  \centering
    \centering
    \includegraphics[width=\linewidth]{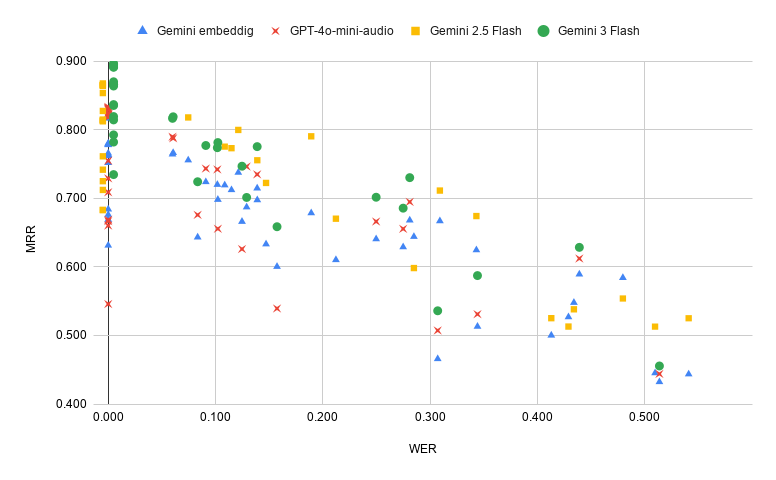}
  \vspace{-2ex}
  \caption{MRR over WER for downstream passage in-lang retrieval in cascaded approach.}
  \label{fig:wer-mrr}
\end{figure}

Furthermore, we analyze the correlation between Word Error Rate (WER) and downstream retrieval performance (e.g., Recall@10) to identify potential information bottlenecks (see Fig.~\ref{fig:wer-recall_at_10}).
First, a clear inverse correlation exists between WER and Recall@10. However, the variance is substantial, with a prediction interval of approximately $0.1$–$0.2$ in Recall@10 matching the spread across languages of the model's performance on transcript truths. 
Notably, while GPT-4o-transcribe consistently outperforms Whisper in terms of WER (indicated by the leftward shift of the GPT-4o-transcribe distribution), Whisper yields better Recall@10 than GPT-4o-transcribe at same WER.
Second, even at remarkably low WER levels (e.g., $5\%$ for en-US), a performance gap persists between the best-performing models and the theoretical upper bound of the ground-truth transcript. Similar trends are observed in the MRR analysis shown in Fig.~\ref{fig:wer-mrr}.

\subsection{Acoustic robustness}
\begin{figure*}[t]
  \centering
  \begin{minipage}{1\linewidth}
    \centering
    \includegraphics[width=.32\linewidth]{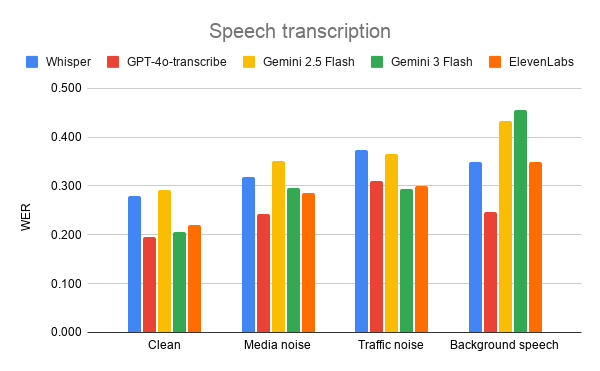}
    \hfill
    \includegraphics[width=.32\linewidth]{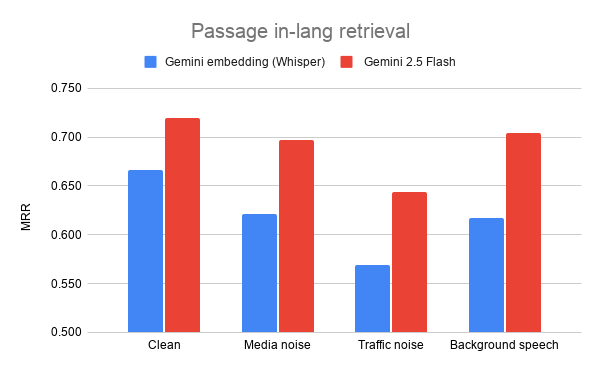}
    \hfill
    \includegraphics[width=.32\linewidth]{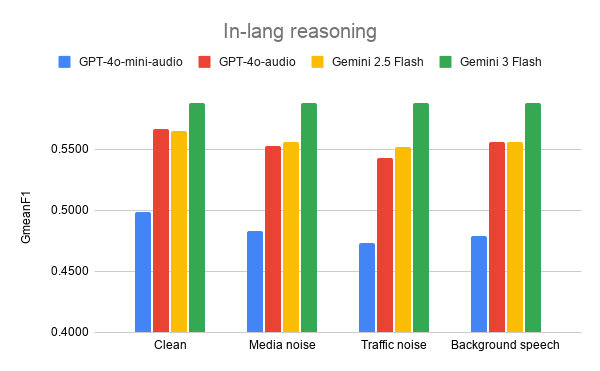}
  \end{minipage}
  \vspace{-2ex}
  \caption{Task performance across different noise conditions.}
  \label{fig:environment-by-task}
\end{figure*}

Fig.~\ref{fig:environment-by-task} provides a breakdown of performance across various environments (clean, media noise, traffic noise, and background speech) for several tasks and models. While models exhibit distinct performance patterns, none perform uniformly across all conditions. A notable exception is reasoning, where the GmeanF1 score remains largely stagnant despite increasing noise. It is unclear whether this task inherently requires less acoustic detail, or if model capacity - rather than environmental noise - is the primary bottleneck, as suggested by the persistently low GmeanF1 scores.

\subsection{Locale parity} 
SVQ includes a diverse set of both high-resource and low-resource locales. Fig.~\ref{fig:wer-scatter} (speech transcription), Fig.~\ref{fig:query-reranking-scatter} (query reranking), and Fig.~\ref{fig:retrieval} (retrieval) illustrate the performance variance across these locales. Overall, there is a substantial performance disparity between the highest and lowest scoring locales, with several locales exhibiting significant performance degradation. This trend persists for both general-purpose LLMs and - arguably to a lesser extent - specialized models, such as GPT-4o-transcribe.


\subsection{Evaluation cost}
Another critical factor in large-scale LLM evaluations is operational overhead, specifically cost and latency. We calculate the price per query using official pricing tiers based on average input and output token counts. To capture the economic spectrum of these models, we consider three use cases: speech transcription (short context, high frequency), reasoning (short context, short response), and document retrieval (long context, high density).
For our calculations, we assume:
\begin{itemize}
    \item \textbf{Speech transcription}: A 4-second audio query (approx. 100 audio tokens) paired with a 100-token text prompt, yielding a 10-token output.
    \item \textbf{Reasoning}: 100 audio tokens paired with a 500-token text prompt, yielding a 100-token output.
    \item \textbf{Document retrieval}: A 40,000-token text context (e.g., a technical manual or legal filing) with a 100-token output response. For simplicity, we ignore the audio tokens.
\end{itemize}

Table~\ref{tab:price} summarizes the pricing for the evaluated models.
\begin{table}[h!]
\centering
\caption{Average price (cent) per query for different APIs.}
\label{tab:price}
\vspace{-2ex}
\begin{tabular}{l|ccc}
\toprule
\textbf{Model} & \textbf{Speech} & \textbf{Reasoning} & \textbf{Document} \\
               & \textbf{transc.} & & \textbf{retrieval} \\
\midrule
GPT-4o-transcribe   & .000004 & & \\
GPT-4o-mini-audio   & .100 & .0135 & .606 \\
GPT-4o-audio        & .435 & .225 & 10.1 \\
Gemini 2.5/3 Flash  & .0155 & .040 / .055 & 1.22 / 2.03 \\
Gemini 3 Pro        & .052 & .220 & 8.12 \\
\bottomrule
\end{tabular}
\end{table}
Given the scale of the SVQ dataset, comprising 177,000 queries per task, the total evaluation cost is significant. 
Regarding performance, GPT-series models generally exhibit higher throughput, often outperforming Gemini models in terms of inference speed by a notable margin (for example, GPT-4o-mini-audio is approximately 10x faster than Gemini 3 Flash).

Specialized models, such as GPT-4o-audio-mini, can achieve high-fidelity speech transcription at a fraction of the cost of general-purpose flagship LLMs. In cascaded pipelines, this offers an economical way to reduce the "audio token" burden by an order of magnitude or more before passing the data to a reasoning model. However, these savings can diminish significantly if the downstream task requires extensive few-shot prompting or results in high-volume output, as the per-token cost of the primary LLM then becomes the dominant factor.

\subsection{Test data contamination}\label{sec:data_contamination}
Test data contamination occurs when information from the intended evaluation set "leaks" into the model's training process, rendering performance metrics unreliable. This is a critical issue in the era of LLMs, as their massive pre-training corpora often inadvertently scrape benchmark data from the web. 

Considering the near-ceiling accuracies for zero-shot intent classification in Table~\ref{tab:llm-by-task}, we suspect Speech-MASSIVE examples permeated the LLM training data. The performance gap in Table~\ref{tab:recall_at_10} is equally telling; Gemini embedding V3's Recall@10 are over 0.1 higher than its peers, strongly suggesting the model encountered data highly correlated with the SVQ data used for the MSEB test suites.
Another notable anomaly emerges in the passage retrieval tasks: the MRR remains above 0.2 even for locales with WER exceeding 100\% (see Fig.~\ref{fig:retrieval} and Fig.~\ref{fig:wer-scatter}). 
This suggests that retrieval success is partially decoupled from transcription verbatim accuracy. While not yet fully investigated, we hypothesize this stems from a combination of hallucinations - where the transcriber adds contextually relevant clutter that the downstream ranker can still utilize - and potential data contamination providing an artificially strong prior.

\section{Conclusion \& Future Work}\label{sec:conclusion}

We demonstrated the application of general-purpose LLMs to MSEB tasks and provided a multi-dimensional evaluation of their performance. 
In addition, this paper includes a rigorous analysis of audio-text parity, demonstrating that a significant modality gap persists across most MSEB tasks. Performance remains heavily contingent on acoustic conditions.
At present, the cascaded and audio-native architectures represent two equally viable but distinct paradigms - the former prioritizing modular control and the latter joint and end-to-end optimization. Ultimately, we argue that regardless of the type and degree of the coupling between speech and language components, these modules must be co-designed and jointly optimized to achieve optimal downstream performance.
Due to computational and budgetary limitations, this paper focuses on the efficiency-oriented variants of the GPT and Gemini families (specifically GPT-4o-mini and Gemini 3 Flash). Future work will expand this evaluation to include high-parameter frontier models and broader model families such as Gemma 3 and Qwen2.5-Omni. Additionally, we intend to incorporate recent multimodal embedding models, such as Amazon Nova and Gemini Multimodal Embeddings, into our RAG-based pipelines. 
We encourage the community to contribute to this open-source effort to build a more comprehensive view of the auditory reasoning landscape.

\bibliographystyle{IEEEtran}
\bibliography{mybib}

\end{document}